\documentclass{WileyMSP-template}
\usepackage{cite}
\usepackage{amssymb}
\usepackage{amsmath}
\usepackage{graphicx}
\usepackage{txfonts}
\usepackage{color}
\usepackage{amsfonts}
\usepackage[colorlinks,urlcolor=blue,linkcolor=blue,citecolor=blue]{hyperref}

\begin{document}

\pagestyle{fancy}
\rhead{\includegraphics[width=2.5cm]{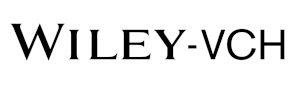}}

\title{Entangling two giant atoms via a topological waveguide}

\maketitle


\author{Wen-Bin Luo}
\author{Xian-Li Yin}
\author{Jie-Qiao Liao\textsuperscript{*}}\\
\begin{affiliations}
Key Laboratory of Low-Dimensional Quantum Structures and Quantum Control of
Ministry of Education, Key Laboratory for Matter Microstructure and Function of Hunan Province, Department of Physics and Synergetic Innovation Center for Quantum Effects and Applications, Hunan Normal University, Changsha 410081, China\\
Email Address:jqliao@hunnu.edu.cn\\
Jie-Qiao Liao\\
Institute of Interdisciplinary Studies Hunan Normal University Changsha 410081, China
\end{affiliations}



\begin{abstract}
The entanglement generation of two two-level giant atoms coupled to a photonic waveguide, which is formed by a Su--Schrieffer--Heeger (SSH) type coupled-cavity array is studied. Here, each atom is coupled to the waveguide through two coupling points. The two-atom separate-coupling case is studied, and 16 coupling configurations are considered for the coupling-point distributions between the two atoms and the waveguide. Quantum master equations are derived to govern the evolution of the two atoms and characterize atomic entanglement by calculating the concurrence of the two-atom states. It is found that the two giant-atom entanglement depends on the coupling configurations and the coupling-point distance of the giant atoms. In particular, the entanglement dynamics of the two giant atoms in 14 coupling configurations depend on the dimerization parameter of the SSH waveguide. According to the self-energies of the two giant atoms, it is found that ten of these 16 coupling configurations can be divided into five pairs. It is also showed that the delayed sudden birth of entanglement between the two giant atoms is largely enhanced in these five pairs of coupling configurations. This work will promote the study of quantum effects and coherent manipulation in giant-atom topological-waveguide-QED systems.
\end{abstract}


\section{Introduction}
Quantum entanglement\textsuperscript{\cite{Schroedigner35,Einstein35,Horodecki09}} is not only a key concept for understanding fundamental quantum theory, but also an important resource for implementing modern quantum technologies, including quantum\\ computation\textsuperscript{\cite{Chuang10}}, quantum communication\textsuperscript{\cite{Bennett92}}, quantum simulator\textsuperscript{\cite{Buluta09}}, quantum precision measurement\textsuperscript{\cite{Giovannetti04,Giovannetti11}}, and quantum sensing\textsuperscript{\cite{Cappellaro17}}. So far, much attention has been paid to the study of quantum entanglement, especially focusing on preparing entangled states in various quantum systems, quantifying quantum entanglement, and exploiting quantum entanglement for practical applications. In particular, both theoretical and experimental studies on quantum entanglement have been conducted in various quantum systems, including optical systems\textsuperscript{\cite{Pan12}}, atom-cavity systems\textsuperscript{\cite{Haroche2001,Rempe2008}}, trapped ion systems\textsuperscript{\cite{Wineland03,Wineland09}}, quantum dot systems\textsuperscript{\cite{Gao2012,Lodahl15}}, Rydberg atom arrays\textsuperscript{\cite{Lukin19}},  superconducting quantum circuits\textsuperscript{\cite{Nori2013,Gu2017,Zhu19,Pan23}}, cavity optomechanical systems\textsuperscript{\cite{Vitali07,Lehnert13,Liao14,Jiao20,Lai22}}, and waveguide quantum electrodynamics (QED) systems\textsuperscript{\cite{Sheremet21,Roy17}}. Waveguide QED addresses the interactions between atoms and photons within a 1D open-boundary waveguide. In a 1D waveguide, photons are restricted to move in the waveguide direction due to the confinement within its cross section. The confinement enhances the interaction between the photons and atoms, enabling quantum entanglement among distant atoms.  Thus, the waveguide-QED systems can provide outstanding platforms to construct various light-matter interactions.  Many interesting physical phenomena have been found in waveguide-QED systems, including single- or few-photon scatterings\textsuperscript{\cite{Shen2005a,Shen2005b,Zhou2008,Tsoi2008,Shi2009,Liao2009,Tsoi2009,Liao2010a,Baragiola2012,Liao2013,ZLiao2016}}, waveguide-mediated long-distance entanglement between distant atoms\textsuperscript{\cite{Vidal11,Baranger13}}, and creating superradiant and subradiant atomic states\textsuperscript{\cite{DeVoe1996,vanLoo2013,Zanner2022}}.

Recently, the giant atom has attracted considerable attention from the peers of quantum optics\textsuperscript{\cite{Kockum20Rev}}. Many interesting physical effects appear when considering the giant atoms coupled to the electromagnetic fields. The key point here is that a giant atom couples to the fields through multiple points, and hence the dipole approximation from a global viewpoint is violated. The interference effect between these coupling points can modulate the collective effects of atoms and effectively affect the atom-field interactions. A series of novel quantum optical phenomena and effects have been found in giant-atom-waveguide systems, such as frequency-dependent lamb shifts and relaxation rates\textsuperscript{\cite{Johansson2014}}, exotic bound states\textsuperscript{\cite{Nori2020,Johansson2020,Wang2020,Tudela2021, Yuan2022,Kwek2023,Jia2023}}, decoherence-free interaction\textsuperscript{\cite{Nori2018,Oliver2020,Ciccarello2020}}, non-Markovian dissipative dynamics\textsuperscript{\cite{Guo17PRA,Delsing19,Longhi2020,Yin22,Lu2023}}, and chiral light-matter interactions\textsuperscript{\cite{Du2022,Kockum2022,Wang2022,Joshi23}}.

Motivated by the recent developments on the giant-atom waveguide QED platforms, people have begun to consider the coupling of giant atoms with topological waveguides\textsuperscript{\cite{Tudela2021,Liu2022,Du2023,Bag23,Gao24}}. In general, the topological phase of matter possesses a natural robustness to environmental perturbations\textsuperscript{\cite{Dusuel2011}}. In waveguide-QED systems, the fields in the waveguide are usually considered the environments of the coupled atoms. Meanwhile, quantum entanglement is inherently sensitive to the environment. Therefore, an interesting question is how the topological waveguide environment affects the entanglement properties of the atoms. Note that the couplings of small atoms with topological waveguides\textsuperscript{\cite{Dusuel2011,Kimble15,Bello2019,Zhang20,Painter21,Dong21,Ciccarello21,Dong22}} have been widely studied, and that the use of topological properties to protect quantum entanglement and quantum coherence has wide potential applications\textsuperscript{\cite{Segev16,Hafezi16,Ozawa19,Wang19,Nie20}}.

\begin{figure*}[t]
\center\includegraphics[width=0.6\textwidth]{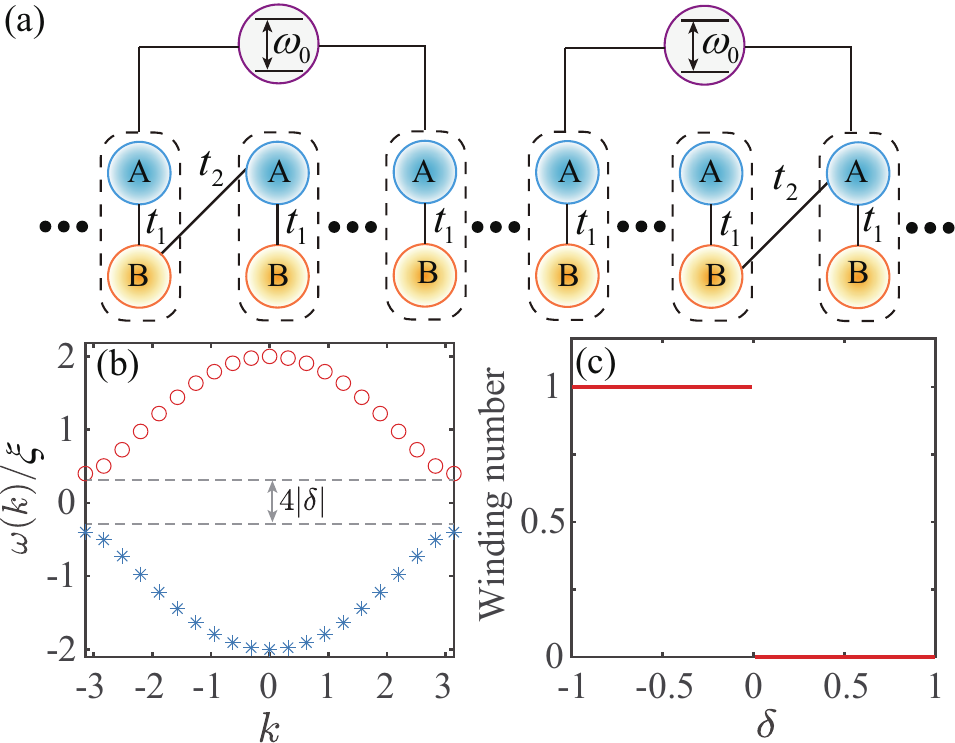}
\caption{a) Schematic of the two two-level giant atoms coupled to a Su-Schrieffer-Heeger type waveguide via $n_{1}$th, $n_{2}$th, $n_{3}$th, and $n_{4}$th unit cells. The left (right) giant atom is labeled as the giant atom $1$ (2). The photonic SSH waveguide includes two different sublattice A and B (blue and yellow circles) with intracell and intercell hopping amplitudes $t_{1}$ and $t_{2}$. The unit cells are indexed by $l\in \lbrack 1,L]$ (form left to right).  b) Dispersion relation of the SSH waveguide with periodical boundary condition. The parameter is set as $\left\vert \delta \right\vert =0.2$.  c) The winding number $W$ as function of the dimerization strength $\delta$. The winding number are either $W=1$, for $\delta<0$ or $W=1$, for $\delta>0$. This defines the topological nontrivial phase ($W=1$) and trivial phase ($W=0$).}
\label{modelandEband}
\end{figure*}

In this paper, we study the generation of quantum entanglement of two two-level giant atoms, which are coupled to a 1D photonic topological waveguide. Here, the 1D topological waveguide is a photonic Su--Schrieffer--Heeger (SSH) type\textsuperscript{\cite{Heeger79}} coupled cavity array (see Figure~\ref{modelandEband}a). Under the Born-Markovian approximation, we eliminate the degrees of freedom of the fields in the waveguide, and obtain the quantum master equation for describing the evolution of the two giant atoms. By calculating the self-energies of the giant atoms at zero temperature, the related coefficients in the quantum master equation can be determined. When the transition frequencies of the giant atoms lie in the band regime (see Figure~\ref{modelandEband}b), we obtain the expressions for the decay rates of the two giant atoms and the exchange interaction strengths between the two giant atoms. We also show that quantum entanglement between the two giant atoms can be generated by coupling the atoms to a topological waveguide. Concretely, we study the entanglement generation for 16 coupling configurations by considering the two giant atoms initially in either the single-excitation state or the two-excitation state. Here, the different coupling configurations arise from the coupling of the two giant atoms with the SSH waveguide at different sublattices. This is distinguished from the recent works concerning the entanglement generation in two giant atoms coupled to a 1D waveguide with linear dispersion relation\textsuperscript{\cite{Santos23,Yin23}}. Furthermore, the coupling configurations of the two giant atoms in Refs.\cite{Santos23,Yin23} are determined by different arrangements of the coupling points. Our results show that the entanglement dynamics of the two giant atoms depends on the coupling configurations and the coupling-point distance. In addition, there exist 14 coupling configurations, where the two-giant-atom entanglement dynamics can be controlled by the sign of the dimerization parameter of the SSH waveguide. According to the self-energies of the giant atoms, ten of these 16 coupling configurations can be divided into five pairs. In the single-excitation initial-state case, the entanglement dynamics have different evolutions for the states $|eg\rangle$ and $|ge\rangle$ since the two giant atoms have unequal frequency shifts and individual decay rates. This feature cannot be found when two separate or braided giant atoms coupled to a 1D  waveguide with a linear dispersion relation. In the two-excitation initial-state case, the entanglement dynamics of the two giant atoms in each pair have the same evolution, and a delayed entanglement sudden birth between the two giant atoms is largely enhanced compared to the other six coupling configurations and small-atom case.

The rest of this paper is organized as follows. In Section~\ref{Physical model}, we introduce the physical model and present the Hamiltonians. In Section~\ref{QMEQ}, we derive quantum master equations for describing the evolution of the two giant atoms in the Markovian regime. In Section~\ref{EqsInrepresention}, we present the equations of motion for the density matrix elements in the eigen representation of the interaction Hamiltonian of the two giant atoms. In Section~\ref{EMofTwoGAs}, we study the quantum entanglement of the two giant atoms corresponding to both the single- and two-excitation initial states by calculating the concurrence of the atomic density matrix. In Section~\ref{Discussion}, we discuss the features of this work, the intrinsic decay of the giant atoms, and the experimental implementation of this scheme. Finally, we present a brief summary of this work in Section~\ref{Conclusions}.

\section{Physical System and Hamiltonian}\label{Physical model}
We consider a giant-atom waveguide-QED system, in which two two-level giant atoms are coupled to a 1D topological waveguide, which is formed by an SSH-type coupled-cavity array. Here, each giant atom is coupled to the waveguide via two coupling points (i.e., two cavities). The Hamiltonian of the total system can be written as
\begin{equation}
\label{total Hamiltonian}
H=H_{S}+H_{B}+H_{\mathrm{int}},
\end{equation}
where $H_{S}$, $H_{B}$, and $H_{\mathrm{int}}$ are the Hamiltonian of the two giant atoms, the Hamiltonian of the waveguide, and the interaction Hamiltonian between the giant atoms and the waveguide, respectively.

Concretely, the Hamiltonian of the two giant atoms reads ($\hbar=1$)
\begin{equation}
\label{HS}
H_{S}=\omega_{0}(\sigma _{1}^{+}\sigma _{1}^{-}+\sigma _{2}^{+}\sigma _{2}^{-}),
\end{equation}
where $\omega_{0}$ is the energy separation between the excited state $|e\rangle_{j=1,2}$ and the ground state $|g\rangle_{j}$ of the $j\mathrm{th}$ giant atom. Note that we set the energy of the ground state $|g\rangle_{j}$ of the atoms as zero in this work. The atomic operators in Equation~(\ref{HS}) are defined by $\sigma_{j=1,2}^{+}=|e\rangle_{jj}\langle g|$ and $\sigma_{j}^{-}=|g\rangle_{jj}\langle e|$, and the Pauli operators are defined by $\sigma_{j}^{x}=\sigma_{j}^{+}+\sigma_{j}^{-}$, $\sigma_{j}^{y}=i(\sigma_{j}^{-}-\sigma_{j}^{+})$, and $\sigma_{j}^{z}=|e\rangle_{jj}\langle e|-|g\rangle_{jj}\langle g|$.

The waveguide under consideration is an SSH-type coupled-cavity array formed by two sublattices, denoted by A and B, as shown in Figure~\ref{modelandEband}a. The Hamiltonian of the SSH-type waveguide reads
\begin{equation}
\label{HB}
H_{B}=\omega_{c} \sum_{l=1}^{L} (a_{l}^{\dag }a_{l}+b_{l}^{\dag }b_{l})+\sum_{l=1}^{L-1}(t_{1}a_{l}^{\dag }b_{l}+t_{2}a_{l+1}^{\dag }b_{l}+\mathrm{H.c.}),
\end{equation}
where $a_{l}$ ($b_{l}$) and $a_{l}^{\dagger}$ ($b_{l}^{\dagger}$) are the annihilation  and creation operators of the cavity $A$ ($B$) in the $l$th unit cell. We assume that all the cavities have the same resonance frequencies $\omega_{c}$, and that the intracell and intercell coupling strengths are, respectively, given by
\begin{equation}
t_{1}=\xi (1+\delta ),\hspace{1cm}t_{2}=\xi (1-\delta ),
\end{equation}
where $\delta $ is introduced as the dimerization parameter, and it satisfies $|\delta|<1$.

To discuss the couplings between the giant atoms and the waveguide, we need to analyze the coupling configurations between the atoms and the cavities in the waveguide. For example, when a giant atom is coupled to the SSH waveguide via two points, there are four coupling configurations: AA, AB, BA, and BB couplings. For the two-atom case, the situation becomes more complicated. When two giant atoms couple to a 1D continuous waveguide, there exist three kinds of coupling configurations: separate, nested, and braided couplings\textsuperscript{\cite{Nori2018}}. In our considered giant-atom-SSH-waveguide system, the coupling configurations become much more complicated. This is because there are two kinds of cavities in a cell unit. To keep compact, in this work we focus on the separate-coupling case, where there are 16 coupling configurations between the atoms and the waveguide: AAAA, AAAB, AABA, AABB, ABAA, ABAB, ABBA, ABBB, BAAA, BAAB, BABA, BABB, BBAA, BBAB, BBBA, and BBBB coupling distributions.

To unify and describe all these coupling configurations, we introduce the parameters $\alpha _{i}$ and $\beta _{i}$ to characterize the couplings between the two giant atoms and the SSH waveguide. The coefficients $\alpha _{i}$ and $\beta _{i}$ take either $0$ or $1$, and satisfy the condition
\begin{equation}
\alpha _{i}+\beta _{i}=1,\hspace{0.5cm}i=1-4.
\end{equation}
For the first coupling point of the giant atom $1$ ($2$) coupled to either the sublattice A or the sublattice B in the same unit cell, the corresponding coupling strengths can be expressed as $g\alpha _{1}$ ($g\alpha_{3}$) or $g\beta _{1}$ ($g\beta _{3}$). Similarly, for the second coupling point of the giant atom $1$ ($2$) coupled to either the sublattice A or the sublattice B in another unit cell, the coupling strengths are denoted as $g\alpha _{2}$ ($g\alpha_{4}$) or $g\beta _{2}$ ($g\beta _{4}$).  If a giant atom is coupled to the cavity A (B) in the $i$th unit cell, we have $\alpha _{i}=1$ ($\beta _{i}=1$). When the coupling is absence, $\alpha _{i}$ ($\beta _{i}$) takes zero. By choosing proper values of $\alpha _{i}$ and $\beta _{i}$, the different coupling configurations can be described in a unified form.

In this work, we consider the weak-coupling regime, where the coupling strength $g$ is much smaller than both the cavity frequency $\omega_{c} $ and the atomic energy separation $\omega_{0}$ (namely $g\ll\{\omega_{c},\omega_{0}\}$). Then the rotating-wave approximation (RWA) can be made by discarding the counter-rotating terms in the interaction Hamiltonian. Based on the above analyses, the interaction Hamiltonian between the two giant atoms and the waveguide can be written as
\begin{eqnarray}
H_{\mathrm{int}} &=&g[(\alpha _{1}a_{n_{1}}+\beta _{1}b_{n_{1}}+\alpha_{2}a_{n_{2}}+\beta _{2}b_{n_{2}})\sigma _{1}^{+}  \nonumber \\
&&+(\alpha _{3}a_{n_{3}}+\beta _{3}b_{n_{3}}+\alpha _{4}a_{n_{4}}+\beta_{4}b_{n_{4}})\sigma _{2}^{+}+\mathrm{H.c.}].
\end{eqnarray}
Here, we assume that the giant atom 1 is coupled to the SSH-chain waveguide at both the $n_{1}$th and $n_{2}$th unit cells, while the giant atom 2 is coupled to the waveguide at both the $n_{3}$th and $n_{4}$th unit cells. All these coupling points have the same coupling strength $g$.

In this system, the SSH-chain waveguide plays the role of an environment of the two giant atoms.  To clearly see the influence of the waveguide on the atoms, we first analyze the field modes in the waveguide by introducing the normal modes to diagonalize the waveguide Hamiltonian, and then we express the atom-waveguide interactions with these normal modes.

For simplicity, we consider the periodic boundary conditions $a_{L+l}=$ $a_{l}$ and $b_{L+l}=$ $b_{l}$. By introducing the discrete Fourier transformations
\begin{equation}
a_{l}=\frac{1}{\sqrt{L}}\sum_{k}e^{ikld_{0}}a_{k},\hspace{0.5cm} b_{l}=\frac{1}{\sqrt{L}}\sum_{k}e^{ikld_{0}}b_{k},
\end{equation}
where $d_{0}$ is the photonic lattice constant (hereafter we take $d_{0}=1$ for simplicity) and $l$ marks different lattice sites. The waveguide Hamiltonian in the momentum space can be written as $H_{B}=\sum_{k}\psi
^{\dag }(k)\mathcal{H}(k)\psi (k)$, with $\psi(k) =\left(a_{k},b_{k}\right) ^{T}$, and the bulk momentum-space Hamiltonian reads
\begin{equation}
\mathcal{H}(k)=\left(
\begin{array}{cc}
\omega_{c} & f(k) \\
f^{\ast }(k) & \omega_{c}
\end{array}
\right).
\end{equation}
Here, $f(k)=\xi (1+\delta )+\xi (1-\delta )e^{-ik}=\omega_{k}e^{i\phi(k)}$ with $\omega _{k}>0$ is the coupling coefficient in the momentum space between A and B sublattices. The topological-dependent phase $\phi(k)$ is given by
\begin{equation}
\label{phik}
\phi (k)=\mathrm{arctan}\left( \frac{-\xi (1-\delta )\sin k}{\xi (1+\delta)+\xi (1-\delta )\cos k}\right) .
\end{equation}
Hereafter we set $\omega_{c}$ as an energy reference. By introducing the eigen-operators
\begin{equation}
u_{k}/l_{k}=\frac{1}{\sqrt{2}}(e^{-i\phi(k) }a_{k}\pm b_{k}),
\end{equation}
the Hamiltonian $H_{B}$ in Equation~(\ref{HB}) can be diagonalized as
\begin{equation}
H_{B}=\sum_{k}[\omega _{u}(k)u_{k}^{\dag }u_{k}+\omega _{l}(k)l_{k}^{\dag}l_{k}]
\end{equation}
with the dispersion relations
\begin{equation}
\omega _{u/l}(k)=\pm \omega _{k}=\pm \sqrt{t_{1}^{2}+t_{2}^{2}+2t_{1}t_{2}\cos k},
\end{equation}
where the subscript $u$ ($l$) denotes the upper (lower) energy band of the SSH waveguide.

In Figure~\ref{modelandEband}b, we show the dispersion relations when the dimerization parameter $|\delta| =0.2$. It can be found that the two bands of the SSH waveguide are symmetric with respect to the cavity frequency $\omega_{c}$. The range of the energy bands is $[-2\xi,-2|\delta |\xi]\cup \lbrack 2|\delta|\xi,2\xi]$, with a middle bandgap $4|\delta|\xi$. The energy bands can be tuned by changing the dimerization strength $\delta$. For the SSH model, we use the winding number $W$ to characterize the topological properties of the system. The winding number $W$ is either $1$ or $0$, depending on the parameters of the system. In Figure~\ref{modelandEband}c, we show the winding number $W$ of the SSH waveguide. When the intracell hopping strength dominates the intercell hopping strength, i.e., $\delta>0$, the winding number $W=0$, corresponds to the so-called topological trivial phase. On the contrary, when the intercell hopping strength dominates the intracell hopping strength, namely $\delta<0$, the winding number $W=1$, corresponding to the so-called topological phase.

For convenience, hereafter we will work in a rotating frame with respect to $H_{0}=\omega _{c}(\sigma _{1}^{+}\sigma _{1}^{-}+\sigma _{2}^{+}\sigma_{2}^{-})+\omega _{c}\sum_{l=1}^{L}( a_{l}^{\dag }a_{l}+b_{l}^{\dag}b_{l}) $, then the Hamiltonian of the system becomes
\begin{eqnarray}
H_{\mathrm{rot}} &=&\Delta (\sigma _{1}^{+}\sigma _{1}^{-}+\sigma_{2}^{+}\sigma _{2}^{-})+\sum_{l=1}^{L}(t_{1}a_{l}^{\dag}b_{l}+t_{2}a_{l+1}^{\dag }b_{l}+\mathrm{H.c.})  \nonumber \\
&&+g[(\alpha _{1}a_{n_{1}}+\beta _{1}b_{n_{1}}+\alpha _{2}a_{n_{2}}+\beta_{2}b_{n_{2}})\sigma _{1}^{+}  \nonumber \\
&&+(\alpha _{3}a_{n_{3}}+\beta _{3}b_{n_{3}}+\alpha _{4}a_{n_{4}}+\beta_{4}b_{n_{4}})\sigma _{2}^{+}+\mathrm{H.c.}].
\end{eqnarray}
Using the discrete Fourier transformations and the eigen-operators, the Hamiltonian $H_{\mathrm{rot}}$ can be expressed in the momentum space as
\begin{eqnarray}
H_{\mathrm{rot}} &=&\Delta (\sigma _{1}^{+}\sigma _{1}^{-}+\sigma_{2}^{+}\sigma _{2}^{-})+\sum\limits_{k}\omega _{k}(u_{k}^{\dag}u_{k}-l_{k}^{\dag }l_{k})  \nonumber \\
&&+\frac{g}{\sqrt{2L}}\sum\limits_{k}[(\alpha _{1}e^{ikn_{1}}+\alpha_{2}e^{ikn_{2}})(u_{k}+l_{k})e^{i\phi (k)}\sigma _{1}^{+}  \nonumber \\
&&+(\beta _{1}e^{ikn_{1}}+\beta _{2}e^{ikn_{2}})(u_{k}-l_{k})\sigma _{1}^{+}\nonumber \\
&&+(\alpha _{3}e^{ikn_{3}}+\alpha _{4}e^{ikn_{4}})(u_{k}+l_{k})e^{i\phi(k)}\sigma _{2}^{+}  \nonumber \\
&&+(\beta _{3}e^{ikn_{3}}+\beta _{4}e^{ikn_{4}})(u_{k}-l_{k})\sigma _{2}^{+}+\mathrm{H.c.}],
\end{eqnarray}
where $\Delta=\omega_{0}-\omega_{c}$ is the detuning between the atomic transition frequency $\omega_{0}$ and the cavity frequency $\omega_{c}$.

\section{Quantum master equation and coefficients}\label{QMEQ}
To govern the evolution of the two giant atoms, we derive a quantum master equation by using the Born-Markovian approximation in the weak-coupling regime $g/\xi\ll 1$\textsuperscript{\cite{ZollerBook}}. The quantum master equation in a rotating frame with respect to $H_{0}$ reads
\begin{eqnarray}
\label{master equation}
\dot{\rho} &=&i\left[ \rho ,\mathcal{H}\right] +\Gamma _{11}\mathcal{D}[\sigma _{1}^{-}]\rho +\Gamma _{22}\mathcal{D}[\sigma _{2}^{-}]\rho\nonumber \\
&&+\Gamma _{12}\left\{ \left[ \sigma _{1}^{-}\rho \sigma _{2}^{+}-\frac{1}{2}(\sigma _{2}^{+}\sigma _{1}^{-}\rho +\rho \sigma _{2}^{+}\sigma _{1}^{-})\right] +\mathrm{H.c.}\right\} ,
\end{eqnarray}
where $\rho $ is the reduced density matrix of the two giant atoms. In Equation~(\ref{master equation}), the Hamiltonian is given by
\begin{equation}
\label{Hamiltonian}
\mathcal{H}=\sum_{j=1}^{2}\omega_{j}^{\prime}\sigma _{j}^{+}\sigma_{j}^{-}+(J_{12}\sigma _{1}^{+}\sigma _{2}^{-}+\mathrm{H.c.}),
\end{equation}
where $\omega_{j}^{\prime}=\Delta+J_{jj}$ (for $j=1$, $2$) with $J_{jj}$ being the frequency shift of the giant atoms. The $J_{12}$ term denotes the exchange interaction between the two giant atoms. The $\Gamma _{11}$ and $\Gamma _{22}$ terms describe the individual decay of the giant atoms $1$ and $2$, respectively,  and the $\Gamma _{12}$ term describes the collective decay of the two giant atoms. $\mathcal{D}[o]\rho\equiv o\rho o^{\dag }-\{o^{\dag}o\rho+\rho o^{\dag}o\}/2$ is the Lindblad superoperator describing the decay of the giant atom.

We note that the variables $J _{ij}$ and $\Gamma_{ij}$ are, respectively, the real and imaginary parts of the collective self-energies $\Sigma _{ij}(\Delta+i0^{+})=J_{ij}-i\Gamma _{ij}/2$. For a two-band model considered in this work, the collective self-energies $\Sigma_{ij}(z) $ are expressed as
\begin{equation}
\label{Dofselfenergy}
\Sigma _{ij}(z) =\sum_{k,}\sum_{\alpha =u/l}\frac{\left\langle0\right\vert \sigma _{j}^{-}H_{\mathrm{int}}\alpha _{k}^{\dag }\left\vert0\right\rangle \left\langle 0\right\vert \alpha _{k}H_{\mathrm{int}}\sigma_{i}^{+}\left\vert 0\right\rangle }{z-\omega _{\alpha }(k)},
\end{equation}
where the state $|0\rangle \equiv|gg\rangle |\emptyset\rangle $ denotes that the two giant atoms are in their ground states and the fields in the SSH waveguide are in the vacuum state $|\emptyset\rangle $. The operators $u_{k}$ and $l_{k}$ are the eigen-operators associated with the upper and lower bands, and $k$ is the wave vector within the first Brillouin zone, i.e., $k\in \left[ -\pi ,\pi \right)$. After some lengthy calculations, the collective self-energies in Equation~(\ref{Dofselfenergy}) can be calculated as follows,
\begin{eqnarray}\label{selfenergy}
\Sigma _{11}(z) &=&\frac{2g^{2}}{L}\sum_{k}\{z+z(\alpha _{1}\alpha_{2}+\beta _{1}\beta _{2})\cos[k( n_{1}-n_{2})]\nonumber \\
&&+\omega _{k}\alpha _{1}\beta _{2}\cos [k(n_{1}-n_{2})+\phi (k)]  \nonumber \\
&&+\omega _{k}\alpha _{2}\beta _{1}\cos [k(n_{1}-n_{2})-\phi (k)]\}(z^{2}-\omega _{k}^{2})^{-1},  \nonumber \\
\Sigma _{12}(z) &=&\frac{g^{2}}{L}\sum_{k}[z(\alpha _{1}e^{-ikn_{1}}+\alpha_{2}e^{-ikn_{2}})(\alpha _{3}e^{ikn_{3}}+\alpha _{4}e^{ikn_{4}})  \nonumber\\
&&+f^{\ast }(k)(\alpha _{1}e^{-ikn_{1}}+\alpha _{2}e^{-ikn_{2}})(\beta_{3}e^{ikn_{3}}+\beta _{4}e^{ikn_{4}})  \nonumber \\
&&+f(k)(\beta _{1}e^{-ikn_{1}}+\beta _{2}e^{-ikn_{2}})(\alpha_{3}e^{ikn_{3}}+\alpha _{4}e^{ikn_{4}})  \nonumber \\
&&+z(\beta _{1}e^{-ikn_{1}}+\beta _{2}e^{-ikn_{2}})(\beta_{3}e^{ikn_{3}}+\beta _{4}e^{ikn_{4}})](z^{2}-\omega _{k}^{2})^{-1},\nonumber \\
\Sigma _{22}(z) &=&\frac{2g^{2}}{L}\sum_{k}\{z+z(\alpha _{3}\alpha_{4}+\beta _{3}\beta _{4})\cos [k(n_{3}-n_{4})]  \nonumber \\
&&+\alpha _{3}\beta _{4}\omega _{k}\cos [k(n_{3}-n_{4})+\phi(k)]\nonumber \\
&&+\alpha _{4}\beta _{3}\omega _{k}\cos [k(n_{3}-n_{4})-\phi(k)]\}(z^{2}-\omega _{k}^{2})^{-1},
\end{eqnarray}
and $\Sigma _{12}^{\ast }=\Sigma _{21}$. Equation~(\ref{selfenergy}) indicates that the self-energies depend on the sign of the dimerization parameter $\delta$ due to $\phi(k)$ appearing as a new phase in the expressions of the self-energies. In this case, quantum interference effect comes from the joint contribution of the  phase $\phi(k)$ and the propagating phase between coupling points. Expanding Equation~(\ref{selfenergy}), we find that there are three types of functions:
\begin{eqnarray}
\label{abc}
A_{n_{ij}}(z) &=&\frac{g^{2}}{L}\sum_{k}\frac{ze^{ikn_{ij}}}{z^{2}-\omega_{k}^{2}},  \nonumber \\
B_{n_{ij}}(z) &=&\frac{g^{2}}{L}\sum\limits_{k}\frac{\omega _{k}e^{i\left[kn_{ij}-\phi(k) \right] }}{z^{2}-\omega ^{2}(k)},\nonumber \\
C_{n_{ij}}(z) &=&\frac{g^{2}}{L}\sum\limits_{k}\frac{\omega _{k}e^{i\left[kn_{ij}+\phi(k) \right] }}{z^{2}-\omega ^{2}(k)}.
\end{eqnarray}
Then, the collective self-energies $\Sigma _{ij}(z)$ of the two giant atoms can be expressed with Equation~(\ref{abc}) as
\begin{eqnarray}
\label{selfenergy2}
\Sigma _{11}(z) &=&2[A_{n_{11}}(z)+(\alpha _{1}\alpha _{2}+\beta _{1}\beta_{2})A_{n_{12}}(z)  \nonumber \\
&&+\alpha _{1}\beta _{2}B_{n_{12}}(z)+\alpha _{2}\beta _{1}C_{n_{12}}(z)],\nonumber \\
\Sigma _{12}(z) &=&(\alpha _{1}\alpha _{3}+\beta _{1}\beta_{3})A_{n_{13}}(z)+(\alpha _{2}\alpha _{3}+\beta _{2}\beta _{3})A_{n_{23}}(z)\nonumber \\
&&+(\alpha _{1}\alpha _{4}+\beta _{1}\beta _{4})A_{n_{14}}(z)+(\alpha_{2}\alpha _{4}+\beta _{2}\beta _{4})A_{n_{24}}(z)  \nonumber \\
&&+\alpha _{1}\beta _{3}B_{n_{13}}(z)+\alpha _{2}\beta_{3}B_{n_{23}}(z)\nonumber \\
&&+\alpha _{1}\beta _{4}B_{n_{14}}(z)+\alpha _{2}\beta_{4}B_{n_{24}}(z)  \nonumber \\
&&+\alpha _{3}\beta _{1}C_{n_{13}}(z)+\alpha _{3}\beta_{2}C_{n_{23}}(z)\nonumber \\
&&+\alpha _{4}\beta _{1}C_{n_{14}}(z)+\alpha _{4}\beta_{2}C_{n_{24}}(z),  \nonumber \\
\Sigma _{22}(z) &=&2[A_{n_{33}}(z)+(\alpha _{3}\alpha _{4}+\beta _{3}\beta_{4})A_{n_{34}}(z)  \nonumber \\
&&+\alpha _{3}\beta _{4}B_{n_{34}}(z)+\alpha _{4}\beta _{3}C_{n_{34}}(z)].
\end{eqnarray}

In the thermodynamic limit $(L\rightarrow \infty )$, the sum $\sum_{k}$ can be turn into an integral $\frac{L}{2\pi }\int dk$ using the Weisskopf-Wigner approximation. Applying the residual theorem and variable substitutions $y\equiv e^{ik}$ for $n_{ij}\geq 0$ and $y\equiv e^{-ik}$ for $n_{ij}<0$, Equation~(\ref{abc}) can be rewritten as
\begin{eqnarray}
\label{coefficients}
A_{n_{ij}}(z) &=&-\frac{g^{2}z[y_{+}^{|n_{ij}|}\Theta_{+}(y_{+})-y_{-}^{|n_{ij}|}\Theta _{-}(y_{+})]}{\sqrt{z^{4}-4\xi ^{2}(1+\delta^{2})z^{2}+16\xi ^{4}\delta ^{2}}},  \nonumber \\
B_{n_{ij}}(z) &=&-\frac{g^{2}\xi[F_{n_{ij}}(y_{+})\Theta_{+}(y_{+})-F_{n_{ij}}(y_{-})\Theta _{-}(y_{+})]}{\sqrt{z^{4}-4\xi ^{2}(1+\delta ^{2})z^{2}+16\xi ^{4}\delta ^{2}}},  \nonumber \\
C_{n_{ij}}(z) &=&-\frac{g^{2}\xi[P_{n_{ij}}(y_{+})\Theta_{+}(y_{+})-P_{n_{ij}}(y_{-})\Theta _{-}(y_{+})]}{\sqrt{z^{4}-4\xi ^{2}(1+\delta ^{2})z^{2}+16\xi ^{4}\delta ^{2}}},
\end{eqnarray}
where $F_{n}(z) =[(1+\delta )z^{|n|}+(1-\delta )z^{|n+1|}]$, $P_{n}(z)=[(1+\delta )z^{|n|}+(1-\delta )z^{|n-1|}]$, and $\Theta_{\pm}(z)=\Theta (\pm 1\mp|z|)$, with $\Theta(z) $ being the unit step function; $n_{ij}=n_{j}-n_{i}$ is the distance between the coupling points $n_{j}$ and $n_{i}$. The two simple poles in Equation~(\ref{coefficients}) are
\begin{equation}
\label{poles}
y_{\pm }=\frac{z^{2}-2\xi ^{2}(1+\delta ^{2}) \pm \sqrt{z^{4}-4\xi ^{2}( 1+\delta ^{2}) z^{2}+16\xi ^{4}\delta ^{2}}}{2\xi ^{2}( 1-\delta ^{2}) }.
\end{equation}
For simplicity, we here assume that the neighboring coupling points of the two giant atoms are equidistant ($n_{4}-n_{3}=n_{3}-n_{2}=n_{2}-n_{1}=d$).

\begin{figure}[t]
\center\includegraphics[width=0.4\textwidth]{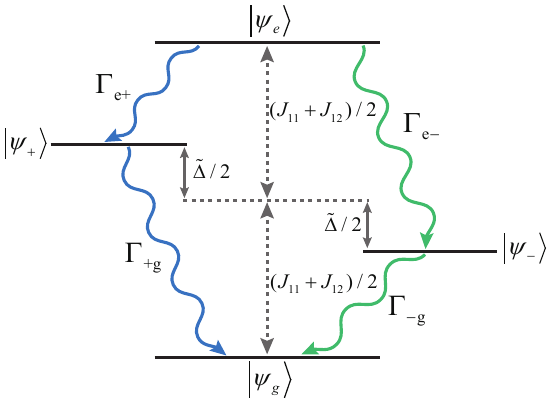}
\caption{Energy levels and transitions in the eigen-representation of the two giant atoms. The parameters $\Gamma_{e\pm}$ ($\Gamma_{\pm g}$) denote the decay rates for the transitions $|\psi_{e}\rangle\rightarrow|\psi_{\pm}\rangle$ ($|\psi_{\pm}\rangle\rightarrow|\psi_{g}\rangle$). The $J_{jj}$ (for $j=1$, $2$) is the Lamb shift of the giant atom $j$ and $\tilde{\Delta}$ is the energy difference between the states $|\psi_{+}\rangle$ and $|\psi_{-}\rangle$.}
\label{energylevel}
\end{figure}

\section{Equations of motion for the density matrix elements in the eigenstate representation of $\mathcal{H}$}\label{EqsInrepresention}
In principle, we can achieve the evolution of the density matrix for the two giant atoms by solving the quantum master equation~(\ref{master equation}). For understanding the evolution of the atoms more clearly, it will be better to work in the eigen-representation of the Hamiltonian $\mathcal{H}$ given in Equation~(\ref{Hamiltonian}). For the Hamiltonian $\mathcal{H}$, the total excitation number operator is $\sigma_{1}^{+}\sigma_{1}^{-}+\sigma_{2}^{+}\sigma_{2}^{-}$, which is a conserved quantity due to $[\sigma_{1}^{+}\sigma_{1}^{-}+\sigma_{2}^{+}\sigma_{2}^{-},\mathcal{H}]=0$. The eigenstates of $\mathcal{H}$ in the zero- and two-excitation subspaces are
\begin{equation}
|\psi _{g}\rangle =|gg\rangle \hspace{0.25cm}\mathrm{and} \hspace{0.25cm}|\psi _{e}\rangle =|ee\rangle,
\end{equation}
respectively, with the corresponding eigenvalues 0 and $\omega_{1}^{\prime}+\omega_{2}^{\prime}$. In the single-excitation subspace, the eigensystem of $\mathcal{H}$ is determined by the eigen-equaiton $\mathcal{H}|\psi_{\pm }\rangle=\lambda _{\pm}|\psi_{\pm }\rangle$. Here, the eigenvalues are given by
\begin{equation}
\lambda_{\pm }=\frac{1}{2}(\omega _{1}^{\prime }+\omega _{2}^{\prime }\pm \tilde{\Delta} )
\end{equation}
with $\tilde{\Delta}=\sqrt{(J_{11}-J_{22})^{2}+4J_{12}^{2}}$ being the energy difference. The corresponding eigenstates take the form as
\begin{equation}
\label{eigenstate}
|\psi_{\pm }\rangle=\mathcal{N}_{\pm }\left(\frac{J_{11}-J _{22}\pm \tilde{\Delta}}{J_{12}}|eg\rangle +2|ge\rangle \right),
\end{equation}
where the normalization constants are defined by
\begin{equation}
\mathcal{N}_{\pm }=\left[\frac{(J_{11}-J_{22}\pm \tilde{\Delta} )^{2}}{J_{12}^{2}}+4\right]^{-1/2}.
\end{equation}

To better exhibit the dynamics of the two atoms, below we work in the eigen-representation of the two-coupled atoms (i.e., in the representation with respect to $\mathcal{H}$). In Figure~\ref{energylevel}, we show the energy levels and transition rates of the two giant atoms in the eigen-representation with the  bases $\{|\psi_{e}\rangle  ,|\psi_{\pm }\rangle,|\psi_{g}\rangle\}$. In this representation, the equations of motion for these density matrix elements are given by
\begin{eqnarray}
\label{EvolutionOfelement}
\dot{\rho}_{ee}(t) &=&-(\Gamma _{11}+\Gamma _{22})\rho _{ee}(t),  \nonumber\\
\dot{\rho}_{++}(t) &=&i\Delta _{1}[\rho _{-+}(t)-\rho _{+-}(t)]+\Gamma_{e+}\rho _{ee}(t)+\Gamma _{++}\rho _{++}(t)  \nonumber \\
&&+\Gamma _{+-}\rho _{+-}(t)+\Gamma _{-+}\rho _{-+}(t),  \nonumber \\
\dot{\rho}_{--}(t) &=&i\Delta _{1}[\rho _{+-}(t)-\rho _{-+}(t)]+\Gamma_{e-}\rho _{ee}(t)+\Gamma _{--}\rho _{--}(t)  \nonumber \\
&&+\Gamma _{+-}\rho _{+-}(t)+\Gamma _{-+}\rho _{-+}(t),  \nonumber \\
\dot{\rho}_{+-}(t) &=&-\left( \frac{\Gamma _{11}+\Gamma _{22}}{2}+i\Delta_{2}\right) \rho _{+-}(t)+(\Gamma _{+-}-i\Delta _{1})\rho _{++}(t)  \nonumber\\
&&+(\Gamma _{-+}+i\Delta _{1})\rho _{--}(t)+\Gamma_{\times}\rho _{ee}(t),\nonumber \\
\dot{\rho}_{-+}(t) &=&-\left( \frac{\Gamma _{11}+\Gamma _{22}}{2}-i\Delta_{2}\right) \rho _{-+}(t)+(\Gamma _{+-}+i\Delta _{1})\rho _{++}(t)  \nonumber\\
&&+(\Gamma _{-+}-i\Delta _{1})\rho _{--}(t)+\Gamma_{\times}\rho _{ee}(t),\nonumber \\
\dot{\rho}_{gg}(t) &=&\Gamma _{+g}\rho _{++}(t)+\Gamma _{-g}\rho_{--}(t)-2\Gamma _{+-}\rho _{+-}(t)-2\Gamma _{-+}\rho _{-+}(t).  \nonumber \\
&&
\end{eqnarray}
where these density matrix elements are defined by $\rho_{\mu\nu}\equiv\langle\psi_{\mu}|\rho|\psi_{\nu}\rangle$ for $\mu$, $\nu=\{e,\pm,g\}$. In addition, we introduce the following parameters
\begin{eqnarray}
\label{transitionrates}
\Gamma _{e\pm } &=&\frac{\pm \eta _{\pm }\Gamma _{22}\mp \eta _{\mp}\Gamma_{11}\pm 4J_{12}\Gamma _{12}}{2\tilde{\Delta}},  \nonumber \\
\Gamma _{\pm g} &=&\frac{\pm \eta _{\pm}\Gamma _{11}\mp \eta _{\mp}\Gamma_{22}\pm 4J_{12}\Gamma _{12}}{2\tilde{\Delta}},  \nonumber \\
\Gamma _{\pm \pm } &=&\frac{\pm \eta _{\mp}\Gamma _{22}\mp \eta _{\pm}\Gamma_{11}\mp 4J_{12}\Gamma _{12}}{2\tilde{\Delta}},  \nonumber \\
\Gamma _{+-} &=&\Gamma _{-+}=\frac{(\Gamma _{11}-\Gamma _{22})\sqrt{-\eta_{+}\eta _{-}}+2J_{12}\Gamma _{12}\zeta}{4\tilde{\Delta}}, \nonumber \\
\Gamma_{\times} &=&\frac{(\Gamma _{11}-\Gamma _{22})\sqrt{-\eta _{+}\eta _{-}}-2J_{12}\Gamma _{12}\zeta}{2\tilde{\Delta}},  \nonumber \\
\Delta _{1} &=&\frac{(J _{11}-J_{22})\sqrt{-\eta _{+}\eta _{-}}+2J_{12}^{2}\zeta}{2\tilde{\Delta}},  \nonumber \\
\Delta _{2} &=&\frac{(J_{11}-J_{22})^{2}+4J_{12}^{2}}{\tilde{\Delta}},
\end{eqnarray}
with
\begin{eqnarray}
\eta _{\pm } &=&J_{11}-J_{22}\pm \tilde{\Delta},  \nonumber \\
\zeta&=&\sqrt{-\frac{\eta _{-}}{\eta _{+}}}-\sqrt{-\frac{\eta _{+}}{\eta _{-}}}.
\end{eqnarray}
The above results indicate that, by changing the coupling configurations, the sign of $\delta$, and the size of the giant atoms, it is possible to modulate the individual decay rates, the Lamb shifts, the collective decay rate, and the exchange interaction between the two giant atoms. Further, the entanglement dynamics of the two giant atoms can be controlled on demand.

\begin{figure}[t]
\center\includegraphics[width=1\textwidth]{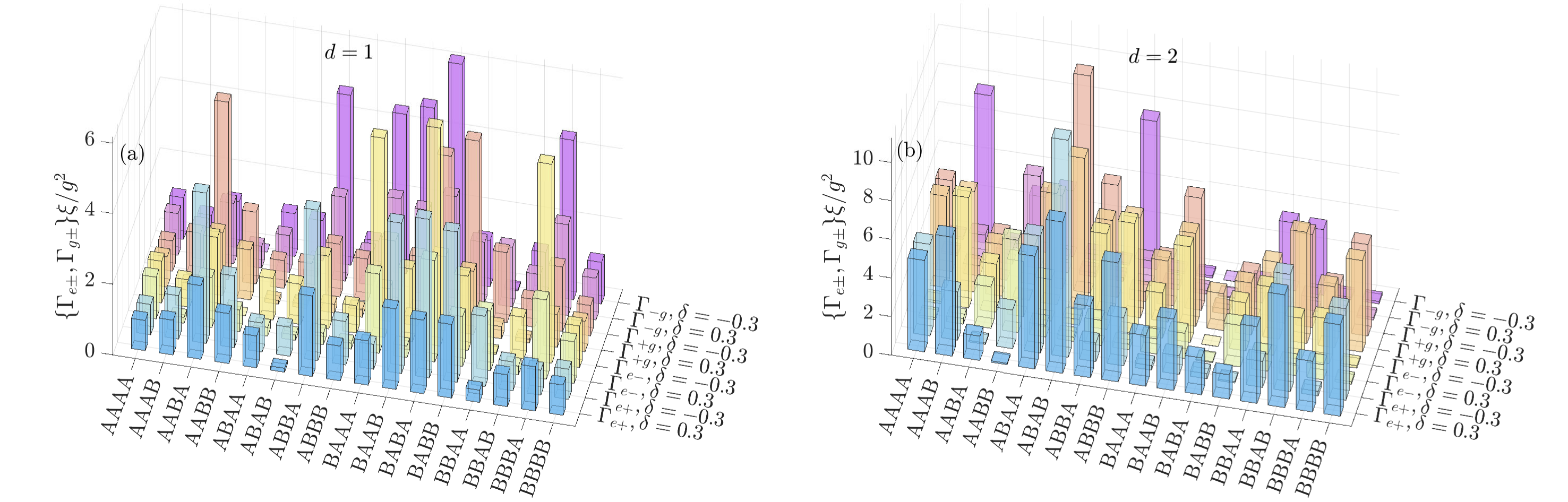}
\caption{Scaled transition rates $\Gamma_{e\pm}\xi/g^{2}$ and $\Gamma_{\pm g}\xi/g^{2}$ for 16 coupling configurations in both the topological trivial phase ($\delta=0.3$) and topological phase ($\delta=-0.3$) of the SSH waveguide. The coupling-point distances $d=1$ and $d=2$ are set for panels  (a) and  (b), respectively. Other parameter used is $\Delta=\xi$.}
\label{bartransitionrates}
\end{figure}

In Figure~\ref{bartransitionrates}a,b, we exhibit the value of these scaled transition rates $\Gamma_{e\pm}\xi/g^{2}$ and $\Gamma_{\pm g}\xi/g^{2}$ corresponding to different coupling configurations and different values of the dimerization parameter $\delta$ when $d=1$ and $d=2$, respectively. For different coupling-point distances, the quantum interference effect induced by the multiple coupling points is different, and hence the transition rates exhibit different features. We find that the four transition rates in the AAAA coupling are equal to those in the BBBB coupling. Meanwhile, the transition rates are immune to the sign of $\delta$ in these two cases. For the remaining 14 couplings, the four transition rates depend on the sign of $\delta$. In addition, for the AAAA, BBBB, ABAB, BABA, AABB, and BBAA couplings, it can be proved that the eigenstates in Equation~(\ref{eigenstate}) are reduced to the symmetric and antisymmetric states. For these six coupling cases, the two giant atoms in Figure~\ref{modelandEband}a have the same frequency shift and the individual decay rate. Therefore, the energy difference becomes $\tilde{\Delta}=2|J_{12}|$ and the parameters $\eta_{\pm}$ are reduced to $\pm2|J_{12}|$. In these six cases, the transition rates in Equation~(\ref{transitionrates}) satisfy the relations $\Gamma_{e\pm}=\Gamma_{\pm g}=\Gamma_{11}\pm J_{12}\Gamma_{12}/|J_{12}|$ when either $\delta>0$ or $\delta<0$, as shown in Figure~\ref{bartransitionrates}. For other ten couplings, the transition rates between the four collective states have different features, exhibiting the generation of larger entanglement compared to the small-atom case.

\section{The entanglement dynamics of the two giant atoms}\label{EMofTwoGAs}
To describe quantum entanglement between the two giant atoms, we adopt concurrence to quantitively measure the entanglement for the density matrix $\rho$ of the two giant atoms. For a two-qubit system, the concurrence is defined by\textsuperscript{\cite{Wootters98}}
\begin{equation}
\label{concurrence}
C(\rho) =\max \left( 0,\lambda _{1}-\lambda _{2}-\lambda_{3}-\lambda _{4}\right),
\end{equation}
where $\lambda _{i}$ (for $i=1$, $2$, $3$, and $4$) are the square root of the eigenvalues of the matrix $\rho \tilde{\rho}$ in decreasing order, i.e., $\lambda_{1}>\lambda _{2}>\lambda _{3}$ $>\lambda _{4}$. The $\rho$ is the density matrix of the two giant atoms, and $\tilde{\rho}\equiv(\sigma_{y}\otimes \sigma _{y}) \rho ^{\ast }(\sigma_{y}\otimes\sigma_{y})$, where the superscript $``*"$ denotes complex conjugation in the bare-state bases \{$|ee\rangle$, $|eg\rangle$, $|ge\rangle$, $|gg\rangle$\}, and $\sigma _{y}$ is the Pauli operator. The concurrence $C(\rho)$ for the two giant atoms satisfies $0\leq C(\rho) \leq 1$. Below, we consider the entanglement generation of the two giant atoms initially in both the single-excitation state $|eg\rangle$ $(|ge\rangle)$ and the two-excitation state $|ee\rangle$.

\begin{figure}[t]
\center\includegraphics[width=0.99\textwidth]{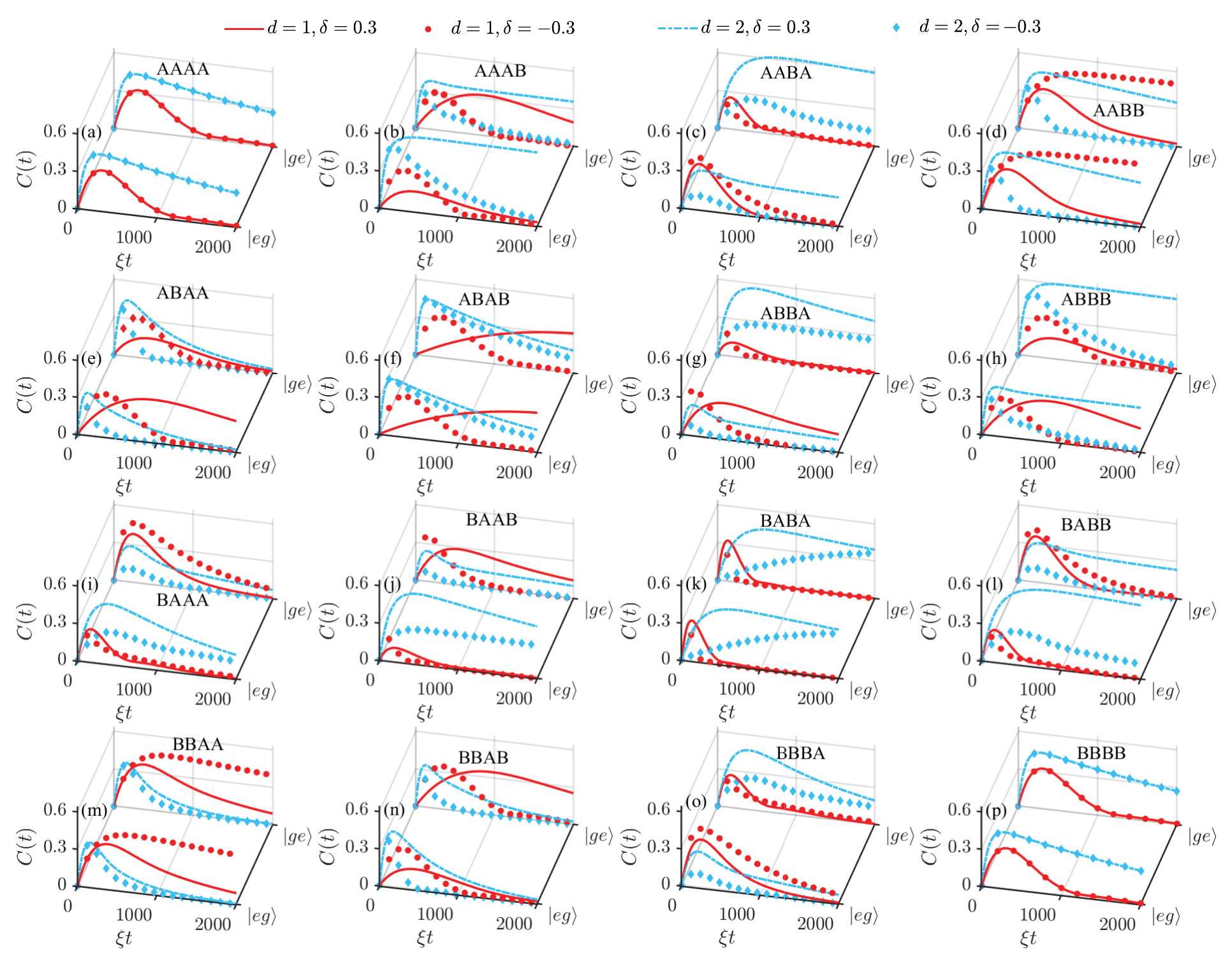}
\caption{Time evolution of the concurrence $C(t)$ of the two giant atoms when the initial state is either $|eg\rangle$ or $|ge\rangle$ corresponding to different coupling configurations:  a) AAAA,  b) AAAB,  c) AABA,  d) AABB,  e) ABAA,  f) ABAB,  g) ABBA,  h) ABBB,  i) BAAA,  j) BAAB,  k) BABA,  l) BABB,  m) BBAA,  n)BBAB,  o) BBBA, and  p) BBBB couplings. Other parameters used are $g=0.05\xi $ and $\Delta=\xi$.}
\label{Cvstegge}
\end{figure}

\subsection{The entanglement of the two giant atoms for the single-excitation initial states $|eg\rangle$ and $|ge\rangle$}
In this subsection, we study the entanglement generation of the two giant atoms for the single-excitation initial states. Since each giant atom is coupled to the waveguide via two coupling cavities, the two giant atoms in some configurations may not satisfy the permutation symmetry. Therefore, we consider these two initial states $|eg\rangle$ and $|ge\rangle$ for the two atoms.  Corresponding to the states $|eg\rangle$ and $|ge\rangle$, the nonzero density-matrix elements are $\langle eg|\rho|eg\rangle= 1$ and $\langle ge|\rho|ge\rangle= 1$, respectively. For the single-excitation initial states, the two-atom system can be understood as a three-level system with bases $\{|\psi_{+}\rangle,|\psi_{-}\rangle,|\psi_{g}\rangle\}$. The coupling of the two atoms with the waveguide will lead to the decay channels $|\psi_{+}\rangle\rightarrow|\psi_{g}\rangle$ and $|\psi_{-}\rangle\rightarrow|\psi_{g}\rangle$. By numerically solving quantum master Equation~(\ref{master equation}) and using Equation~(\ref{concurrence}), the time evolution of the concurrence for the two giant atoms can be obtained. Note that the SSH waveguide is a highly structured bath, and hence the Born--Markovian approximation does not hold near the band edges. To ensure the validity of Equation~(\ref{master equation}), we consider the case where the atomic level $\Delta$ is inside the upper band and $g\ll\xi$.

In Figure~\ref{Cvstegge}, we plot the concurrence $C(t)$ of  the two giant atoms as a function of the scaled time $\xi t$ under different initial states for 16 coupling configurations. The red solid curve and red dots correspond to the parameters $(d=1$, $\delta =0.3)$ and $(d=1$, $\delta =-0.3)$, respectively. The blue solid curve and blue rhombuses correspond to $(d=2$, $\delta =0.3)$ and $(d=2$, $\delta =-0.3)$, respectively. Figure~\ref{Cvstegge} shows that the entanglement generation depends on the coupling configuration, the coupling distance, the atomic initial state, and the dimerization parameter $\delta$.

For both the AAAA and BBBB coupling configurations, due to the permutation symmetry of the two giant atoms, the concurrences in Figure~\ref{Cvstegge}\textcolor{blue}{a,p} are characterized by the same evolution when the two giant atoms are initially in either $|eg\rangle$ or $|ge\rangle$. In particular, the concurrences in these two coupling configurations do not change when the sign of $\delta$ is modulated such that the waveguide enters different topological phases. This means that the concurrences are independent of the dimerization parameter in these two cases. As shown in Figure~\ref{Cvstegge}\textcolor{blue}{a,p}, the markers (for $\delta<0$) overlap with the solid curves (for $\delta>0$). To explain this feature, we substitute $\alpha_{1}=\alpha_{2}=\alpha_{3}=\alpha_{4}=1$ (for the AAAA-coupling case) or $\beta_{1}=\beta_{2}=\beta_{3}=\beta_{4}=1$ (for the BBBB-coupling case) into Equation~(\ref{selfenergy}) and find that $\Sigma_{11}^{\mathrm{AAAA}}=\Sigma_{11}^{\mathrm{BBBB}}=\Sigma_{22}^{\mathrm{AAAA}}=\Sigma_{22}^{\mathrm{BBBB}}$ and $\Sigma_{12}^{\mathrm{AAAA}}=\Sigma_{12}^{\mathrm{BBBB}}$. In addition, the phase $\phi(k)$ terms in Equation~(\ref{selfenergy}) are equal to zero. According to the relation $\Sigma_{ij}=J_{ij}-i\Gamma_{ij}/2$, we know that the coefficients $J_{11}$, $J_{22}$, $J_{12}$, $\Gamma_{11}$, $\Gamma_{22}$, and $\Gamma_{12}$ are independent of the phase $\phi(k)$ [see Equation~(\ref{phik})], and hence the concurrences in the AAAA- and BBBB-couplings cannot be modulated by the dimerization parameter $\delta$.

Figure~\ref{Cvstegge}\textcolor{blue}{b-o} show the time evolution of $C(t)$ for other 14 coupling configurations. In these cases, the phase $\phi(k)$ works in Equation~(\ref{selfenergy}). Thus, the concurrences depend on the sign of $\delta$ when the two giant atoms are in the initial states $|eg\rangle$ or $|ge\rangle$. For most coupling configurations, the concurrences decay to zero as the time increases.  In particular, for the AAAB, AABA, AABB, ABBB, and BBAA coupling configurations, the concurrence decays very slowly under some parameter conditions. For example, in the AABB-coupling case, the concurrences for the initial states $|eg\rangle$ and $|ge\rangle$ maintain mostly $\approx0.5$ in the concerned timescale, as shown by the red dots in Figure~\ref{Cvstegge}d. To explain this feature, we substitute $d=1$ and $\delta=0.3$ into Equation~(\ref{transitionrates}) and obtain $\Gamma_{-g}\xi/g^{2}\approx 1.3\times10^{-4}$. This result is consistent with that in Figure~\ref{bartransitionrates}a. In this case, the population of the state $|\psi_{-}\rangle$ can maintain its initial value over a long timescale. This feature can also be observed in the AAAB- and ABBB-coupling configurations, as shown by the blue dashed curves in Figure~\ref{Cvstegge}\textcolor{blue}{b,h}.  Therefore, by properly choosing the parameters and the coupling configurations, we can generate a relatively steady entanglement between the two giant atoms on a relatively long timescale.

We now analyze the influence of the initial states $|eg\rangle$ and $|ge\rangle$ on the entanglement dynamics in detail. Figure~\ref{Cvstegge}\textcolor{blue}{a,d,f,k,m,p} shows that for the AAAA, AABB, ABAB, BABA, BBAA, and BBBB coupling configurations, the concurrence of the two giant atoms in each coupling configuration has the same evolution features for the initial states $|eg\rangle$ and $|ge\rangle$. This is because the two giant atoms have equal frequency shifts and individual decay rates in these six coupling configurations, and the states $|\psi_{\pm}\rangle$ are reduced to the symmetric and antisymmetric states, as discussed in Section~\ref{EqsInrepresention}. Meanwhile, the transition rate $\Gamma_{+-}$ and the parameters $\Gamma_{\times}$, $\Delta_{1}$, $\Delta_{2}$ are equal to zero. Therefore, Equation~(\ref{EvolutionOfelement}) is reduced to
\begin{eqnarray}
\label{EvolutionOfelement2}
\dot{\rho}_{ee}(t) &=&-(\Gamma _{11}+\Gamma _{22})\rho _{ee}(t),  \nonumber\\
\dot{\rho}_{++}(t) &=&\Gamma _{e+}\rho _{ee}(t)+\Gamma _{++}\rho _{++}(t),\nonumber \\
\dot{\rho}_{--}(t) &=&\Gamma _{e-}\rho _{ee}(t)+\Gamma _{--}\rho _{--}(t),\nonumber \\
\dot{\rho}_{gg}(t) &=&\Gamma _{+g}\rho _{++}(t)+\Gamma _{-g}\rho _{--}(t).
\end{eqnarray}
The coefficients $\Gamma_{e\pm}$ and $\Gamma_{\pm g}$ in Equation~(\ref{EvolutionOfelement2}) are given by Equation~(\ref{transitionrates}), and $\Gamma_{jj}$ (for $j=1$, $2$) is the individual decay rate of the giant atom $j$.

 According to Equation~(\ref{EvolutionOfelement2}), the populations of the states $|\psi_{\pm}\rangle$ for any one of these six coupling configurations are characterized by the same evolution. However, for other ten coupling configurations, the populations of the states $|\psi_{\pm}\rangle$ are governed by Equation~(\ref{EvolutionOfelement}). Thus, the concurrences corresponding to the initial states $|eg\rangle$ and $|ge\rangle$ exhibit different features. From Figure~\ref{Cvstegge}\textcolor{blue}{b,h,c,l,e,n,g,j,i,o}, we find that the concurrences in other ten coupling configurations satisfy the relation
\begin{equation}
C_{eg(ge)}^{O_{1}O_{2}O_{3}O_{4}}(t)=C_{ge(eg)}^{\bar{O}_{4}\bar{O}_{3}\bar{O}_{2}\bar{O}_{1}}(t),
\end{equation}
where the superscript denotes the coupling configuration with $O_{i}\in\{A,B\}$ (for $i=1,2,3,4$), and the $\bar{O_{i}}$ is the complement of $O_{i}$ (i.e., $\bar{A}=B$ and $\bar{B}=A$). The subscript $eg$ $(ge)$ marks the initial atomic state $|eg\rangle$ ($|ge\rangle$). Next, we discuss the AAAB- and ABBB-coupling cases as an example to explain this phenomenon, and other four pairs of coupling configurations share the same characteristics. For the AAAB- and ABBB-coupling configurations, it can be proved that there exist the relations $\Sigma_{11}^{\mathrm{AAAB}}(z)=\Sigma_{22}^{\mathrm{ABBB}}(z)$, $\Sigma_{22}^{\mathrm{AAAB}}(z)=\Sigma_{11}^{\mathrm{ABBB}}(z)$, and $\Sigma_{12}^{\mathrm{AAAB}}(z)=\Sigma_{12}^{\mathrm{ABBB}}(z)$, which result in $J_{11(22)}^{\mathrm{AAAB}}=J_{22(11)}^{\mathrm{ABBB}}$, $\Gamma_{11(22)}^{\mathrm{AAAB}}=\Gamma_{22(11)}^{\mathrm{ABBB}}$, $J_{12}^{\mathrm{AAAB}}=J_{12}^{\mathrm{ABBB}}$, and $\Gamma_{12}^{\mathrm{AAAB}}=\Gamma_{12}^{\mathrm{ABBB}}$. Then the giant atom 1 (2) in the AAAB coupling configuration is equivalent to the giant atom 2 (1) in the ABBB coupling configuration. Therefore, in these five pairs of coupling configurations, the entanglement dynamics of the two giant atoms in the initial state $|eg\rangle$ for the $O_{1}O_{2}O_{3}O_{4}$ coupling configuration is the same as that in the state $|ge\rangle$ for the $\bar{O}_{4}\bar{O}_{3}\bar{O}_{2}\bar{O}_{1}$ coupling configuration.

Since the coefficients $A_{n_{ij}}(z)$, $B_{n_{ij}}(z)$, and $C_{n_{ij}}(z)$ in the self-energies [Equation~(\ref{selfenergy2})] depend on the coupling-point distance $d$, the self-energies $\Sigma_{ij}(z)$ can be adjusted by changing the parameter $d$, and thereby the frequency shift $J_{jj}$, the individual decay rate $\Gamma_{jj}$, the collective decay rate $\Gamma_{12}$, and the exchanging interaction $J_{12}$. Figure~\ref{Cvstegge} shows that the evolution of the concurrence is sensitive to the parameter $d$. For example, when we take $d=2$ and $\delta=0.3$, the concurrence for the AABA coupling configuration can exceed 0.5 and then decays to zero very slowly (the blue dot-dashed curve for the state $|ge\rangle$ in Figure~\ref{Cvstegge}c). However, when we take $d=1$ and $\delta=0.3$, the concurrence can only reach near 0.3 and then decreases very fast (the blue dots for the state $|ge\rangle$ in Figure~\ref{Cvstegge}c). Thus, the coupling-point distance also plays an important role in the entanglement generation of the two giant atoms.

\begin{figure}[t!]
\center\includegraphics[width=0.98\textwidth]{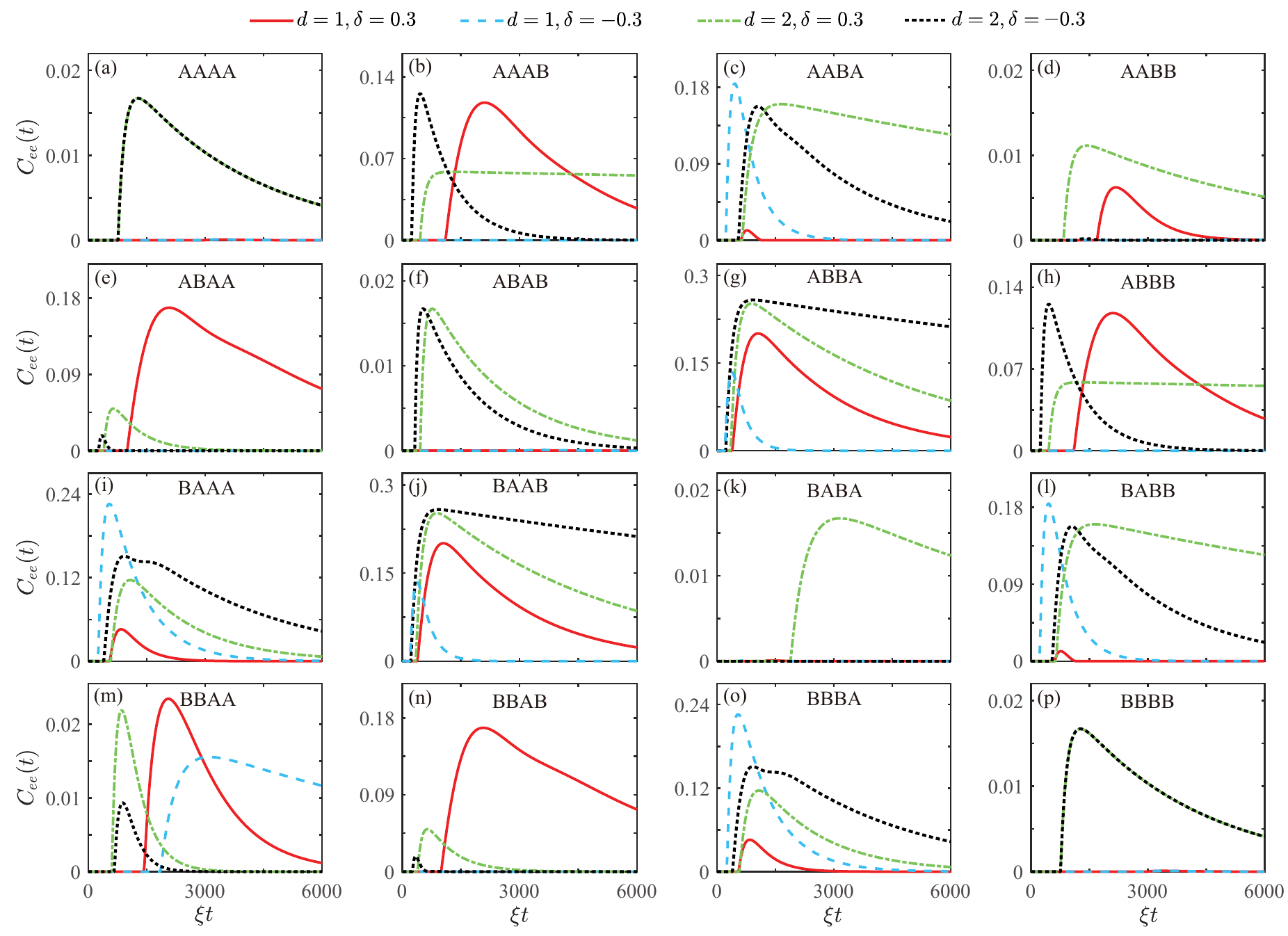}
\caption{Concurrence $C_{ee}(t)$ of the two giant atoms initially in the two-excitation state $|ee\rangle$ as a function of $\xi t$ when the coupling-point distance $d$ and the dimerization parameter $\delta$ take different values. The coupling configurations are:  a) AAAA,  b) AAAB,  c) AABA,  d) AABB,  e) ABAA,  f) ABAB,  g) ABBA,  h) ABBB,  i) BAAA,  j) BAAB,  k) BABA,  l) BABB,  m) BBAA,  n)BBAB,  o) BBBA, and  p) BBBB couplings. Other parameters used are $g=0.05\xi $ and $\Delta=\xi $.}
\label{Cvstee}
\end{figure}

\subsection{The entanglement of the two giant atoms for the two-excitation initial state $|ee\rangle$}
In this subsection, we study the entanglement generation of the two giant atoms initially in the two-excitation state $|ee\rangle$. Different from the single-excitation initial-state case, the dynamics of the two giant atoms in this case can be understood as a four-level system, with two different decay channels. Due to the coupling with the SSH waveguide, the two giant atoms first decay from the initial state $|ee\rangle$ to a mixture of the states $|\psi_{+}\rangle$ and $|\psi_{-}\rangle$, with the decay rates $\Gamma_{e+}$ and $\Gamma_{e-}$, respectively. This leads to the increase of the populations of these two states from zero values. After the populations reach their maximal values, they decay gradually to the ground state $|gg\rangle$, with the corresponding decay rates $\Gamma_{+g}$ and $\Gamma_{-g}$, respectively. In order to generate quantum entanglement of the two giant atoms corresponding to the initial state $|ee\rangle$, we need to choose proper parameters such that the two decay channels are asymmetric.

To study the dependence of the two-atom entanglement on the coupling configurations, coupling-point distance, and the dimerization parameter, we show in Figure~\ref{Cvstee} the dynamics of the concurrence $C(t)$ for these sixteen coupling configurations when $g=0.05\xi$ and $\Delta=\xi$. Different from the single-excitation initial-state case, the two-atom entanglement is generated at later times for the initial state $|ee\rangle$. This phenomenon is known as the delayed sudden birth of entanglement\textsuperscript{\cite{Ficek08,Retamal08}}, which is opposite to the sudden death of entanglement. At earlier times, there is no entanglement generation between the two giant atoms, but at some finite times atomic entanglement builds up suddenly.

We now analyze the dependence of the entanglement generation on the dimerization parameter $\delta$ and coupling configurations in detail. Figure~\ref{Cvstee} shows that, except for the AAAA (Figure~\ref{Cvstee}a) and BBBB (Figure~\ref{Cvstee}p) coupling configurations, the entanglement dynamics for other 14 coupling configurations can be modulated by the sign of $\delta$. This feature is consistent with the case of single-excitation initial states. Figure~\ref{Cvstee}\textcolor{blue}{a,d,f,k,m,p} shows that, for the AAAA, AABB, ABAB, BABA, BBAA, and BBBB coupling configurations, the maximally achievable entanglement is $\approx$0.03, which is negligibly small. It has been proved that the maximally generated entanglement of the two separate or braided giant atoms coupled to a 1D waveguide with linear dispersion relation is also $\approx$0.03\textsuperscript{\cite{Santos23,Yin23}}. To realize a larger entanglement for the two-excitation initial state, the two giant atoms needs to be coupled with the waveguide through the nested coupling configuration.

In our scheme, to realize a larger entanglement for the initial state $|ee\rangle$, the four transition rates $\Gamma_{e\pm}$ and $\Gamma_{\pm g}$ should exhibit different behaviors. For this purpose, we consider that the two giant atoms couple to the waveguide through other ten coupling configurations: AAAB, AABA, ABAA, ABBA, ABBB, ABBB, BAAA, BAAB, BABB,  BBAB, and BBBA. In these ten coupling configurations, we can realize the relations $\Gamma_{+g}>\Gamma_{e+}>\Gamma_{e-}\gg\Gamma_{-g}$ or $\Gamma_{-g}>\Gamma_{e-}>\Gamma_{e+}\gg\Gamma_{+g}$, and then a larger entanglement can be generated. For example, the concurrence in the ABBA-coupling case can exceed 0.2 when $d=2$ and $\delta=\pm 0.3$.  For the AAAB- and ABBB-coupling configurations, the transition rates from the collective states $|\psi_{\pm}\rangle$ to the ground state  $|\psi_{g}\rangle$ satisfy the relation $\Gamma_{-g}\ll\Gamma_{+g}<\xi$. This leads to a slow decay of the population of the state $|\psi_{-}\rangle$. Therefore, in these two coupling cases, the entanglement generation arises from the population of the state $|\psi_{-}\rangle$. Though the transition rate $\Gamma_{-g}$ is much smaller than $\xi$, it is not exactly equal to zero and hence the state $|\psi_{-}\rangle$ is not a dark state. However, within the concerned timescale, the concurrence almost maintains its maximal value, as shown by the green dashed curves in Figure~\ref{Cvstee}\textcolor{blue}{b,h}. According to these analyses, we know that the optimal coupling configuration can be chosen under the given system parameters.  In addition, we would like to point out that the optimally generated entanglement does not always appear in the topological nontrivial phase ($\delta<0$), it can also exist in the topological trivial phase ($\delta>0$), which depends on both the coupling configurations and the coupling-point distance.

By comparing the concurrences of the two giant atoms for the ten coupling configurations in Figure~\ref{Cvstee}\textcolor{blue}{b,h,}\\ \textcolor{blue}{c,l,e,n,g,j,i,o}, it is straightforward to find the relation:
\begin{equation}
C_{ee}^{O_{1}O_{2}O_{3}O_{4}}(t)=C_{ee}^{\bar{O}_{4}\bar{O}_{3}\bar{O}_{2}\bar{O}_{1}}(t),
\end{equation}
where the subscript denotes the initial state of the two giant atoms. In this case, we can also divide these ten coupling configurations into five pairs, similar to the single-excitation initial-state case. The correspondence relation can also be confirmed from the panel pairs: (b)$\leftrightarrow$(h), (c)$\leftrightarrow$(l), (e)$\leftrightarrow$(n), (g)$\leftrightarrow$(j), and (i)$\leftrightarrow$(o).

For the two-excitation initial state $|ee\rangle$, we discuss the influence of the coupling-point distance $d$ on the entanglement generation by taking some of the 16 coupling configurations as examples.  As seen in the AAAA coupling configuration, the concurrence only increases slightly when $d$ increases from 1 to 2.  However, for the ABBA- and BAAB-coupling cases,  the maximal value of the concurrence  increases from 0.13 to 0.26. For the ABAA- and BBAB-coupling cases, the maximal value of the concurrence decreases for a larger $d$. These results mean that the concurrence has a different dependence on $d$ for different coupling configurations and different signs of $\delta$.

\section{Discussion}\label{Discussion}
We now present some discussions to clarify the differences between this work and previously published works\textsuperscript{\cite{Yin22,Santos23,Yin23}}. Concretely, the physical model, motivation, and results in this work are different from those in the two works\textsuperscript{\cite{Yin22,Santos23,Yin23}}. Here, we study the couplings of two separate giant atoms with a structured topological SSH-chain waveguide, which has a different dispersion relation from the linear waveguides considered in Refs.\cite{Yin22,Santos23,Yin23}. In addition, to generate maximally long-lived entangled states, we could consider that the system is driven by a resonant classical field\textsuperscript{\cite{Santos23}} . Except for investigating the influence of the coupling configurations, the coupling distance, and the atomic initial state on entanglement generation, we analyze the influence of the topological-dependent phase $\phi(k)$ on the entanglement dynamics. The detailed analyses on the entanglement generation in different coupling configurations and parameters could provide a guideline for experimental studies on the giant-atom topological-waveguide-QED systems.

We want to point out that one of the original research motivations is to find a relationship between the giant-atom entanglement and the topological properties of the waveguide. In particular, we expect that the relationship can exhibit the same feature as the dependence of the winding number on the dimerization parameter. Namely, when the SSH waveguide transits from the trivial phase to the topological phase, the giant-atom entanglement will experience a sudden change. However, the perturbative treatment of the open system within the Wigner--Weisskopf framework will erase the effect induced by the band gap or band edges. Therefore, in this paper, we cannot obtain a simple expression describing the relationship between the generated entanglement and the topological parameter. Nevertheless, our results show that the topological-dependent phase $\phi(k)$ will affect the entanglement dynamics as an additional phase shift for some coupling configurations.  Recently, the nonperturbative treatment has been adapted to deal with the coupling of small atoms with an SSH waveguide\textsuperscript{\cite{Bello2019}}, where some unconventional quantum optical phenomena are predicted when the atomic transition frequency lies within the bandgap regime of the SSH waveguide.

In this work, the intrinsic decay of the giant atoms is neglected,  because this decay rate is much smaller than the radiative decay rate in typical experiments\textsuperscript{\cite{Oliver2020}}. However, when considering the intrinsic decay, the entanglement generation of the two giant atoms will be suppressed. The larger values of the intrinsic decay rate result in reduced maximum entanglement between the two giant atoms and cause the concurrence to decrease more rapidly to zero. In future research, it remains an interesting topic to study the entanglement dynamics of two giant atoms when their resonance frequencies lie within the band gap. In addition, our scheme can also be extended to a trimer chain or generalized SSH model\textsuperscript{\cite{Dong21}}, which are good platforms to analyze the relationship between the giant-atom entanglement and the topological properties of the structured environment.

Finally, we present some discussions concerning the experimental implementation  of our scheme. The giant atoms have been realized in different platforms, such as transmon qubits coupled to surface acoustic waves\textsuperscript{\cite{Delsing19,Gustafsson14,Manent17}}, Xmon qubits\textsuperscript{\cite{Oliver2020}} or ferromagnetic spin ensembles\textsuperscript{\cite{You22}} coupled to transmission lines. In our scheme, we consider that the giant atom is implemented by a superconducting qubit and the SSH waveguide is implemented by the coupled LC resonator arrays. In this platform, the realization of small atoms coupled to LC resonator arrays has recently been reported\textsuperscript{\cite{Painter21,Houck17,Houck19,Mirhosseini19}}, including the coupling of small atoms with the SSH-type coupled cavity array\textsuperscript{\cite{Painter21}}. Note that the coupling strength between neighboring LC resonators can be adjusted experimentally via periodical modulations\textsuperscript{\cite{Goren18}}, and hence the SSH waveguide can be tuned from the trivial phase to the topological phase by changing the coupling between different resonators. All these advances confirm the experimental feasibility of this scheme with current and near-future conditions.

\section{Conclusion}\label{Conclusions}
In conclusion, we have studied the entanglement dynamics of two two-level giant atoms coupled to a 1D photonic SSH waveguide. We have focused on the two-atom separate coupling configurations, and considered 16 coupling configurations between the two giant atoms and the waveguide. We have also derived a quantum master equation to govern the evolution of the two giant atoms. The dissipation coefficients and the dipole--dipole interaction strength in the master equation have been obtained by calculating the self-energies of the two giant atoms. Based on the quantum master equation, we have characterized the entanglement dynamics of the two giant atoms by calculating the concurrence of the two-atom states for three different atomic initial states: the single-excitation states $|ge\rangle$ and $|eg\rangle$, and the two-excitation state $|ee\rangle$. Except for the AAAA and BBBB coupling configurations, the entanglement dynamics of the two giant atoms in the other 14 coupling configurations depend on the dimerization parameter of the SSH waveguide. By comparing the self-energies of the two giant atoms for the 16 coupling configurations, we have found that ten of these 16 coupling configurations can be divided into five pairs. It is shown that the entanglement dynamics exhibits different features for the above five pairs of coupling configurations in the initial states $|ge\rangle$ and $|eg\rangle$. In addition, we also achieved long-lived entanglement by adjusting the decay process of the two-atom system. In the two-excitation initial state, we have shown that, for these five pairs of coupling configurations, the maximally achievable entanglement is largely enhanced compared to the small-atom scheme and the other six coupling configurations. This work will pave the way for the study of quantum effects and quantum manipulation in giant-atom topological-waveguide-QED systems.

\section*{Acknowledgments}
W.-B. L. and X.-L. Y. contributed equally to this work. J.-Q.L. was supported in part by the National Natural Science Foundation of China (Grants No.~12175061, No.~12247105, and No.~11935006), the Science and Technology Innovation Program of Hunan Province (Grant No.~2021RC4029), and the Hunan Provincial Major Science and Technology Program (Grant No.~2023ZJ1010).  X.-L.Y. was supported in part by the Hunan Provincial Postgraduate Research and Innovation project (Grant No.~CX20230463).

\end{document}